\documentclass[aps, oneprint, twoside, superscriptaddress, nofootinbib,
floatfix, notitlepage, twocolumn]{revtex4-1}

\usepackage{amsmath,amsthm}
\usepackage{amssymb,enumerate}
\usepackage{bm}
\usepackage[dvipsnames]{xcolor}
\usepackage{float}
\usepackage[T1]{fontenc}
\usepackage[pdftex]{graphicx}
\usepackage[pdftex,colorlinks=true,linkcolor=blue,citecolor=blue,urlcolor=blue]{hyperref}
\usepackage{comment}
\usepackage{multirow}
\usepackage{tabularx, booktabs}
\usepackage[normalem]{ulem}
\usepackage{tikz}
\usepackage{yquant}
\usepackage{bm}
\usepackage{pifont}
\usepackage{array}
\usepackage{caption}
\usepackage{subcaption}

\newcommand{\cmark}{\ding{51}}%
\newcommand{\xmark}{\ding{55}}%

\newcommand{\ket}[1]{| #1 \rangle}
\newcommand{\braket}[1]{\langle #1 \rangle}

\newcolumntype{Y}{>{\centering\arraybackslash}X}

\usepackage{booktabs}
\newcommand\headercell[1]{%
   \smash[b]{\begin{tabular}[t]{@{}c@{}} #1 \end{tabular}}}

\begin{document}

\title{Accelerating variational quantum Monte Carlo using the variational quantum eigensolver}

\author{Ashley Montanaro}
\affiliation{Phasecraft Ltd.}
\author{Stasja Stanisic}
\affiliation{Phasecraft Ltd.}

\date{\today}

\begin{abstract}
Variational Monte Carlo (VMC) methods are used to sample classically from distributions corresponding to quantum states which have an efficient classical description. VMC methods are based on performing a number of steps of a Markov chain starting with samples from a simple initial distribution. Here we propose replacing this initial distribution with samples produced using a quantum computer, for example using the variational quantum eigensolver (VQE). We show that, based on the use of initial distributions generated by numerical simulations and by experiments on quantum hardware, convergence to the target distribution can be accelerated compared with classical samples; the energy can be reduced compared with the energy of the state produced by VQE; and VQE states produced by small quantum computers can be used to accelerate large instances of VMC. Quantum-enhanced VMC makes minimal requirements of the quantum computer and offers the prospect of accelerating classical methods using noisy samples from near-term quantum computers which are not yet able to accurately represent ground states of complex quantum systems.
\end{abstract}

\maketitle

The variational quantum eigensolver (VQE) is a quantum algorithm which attempts to produce the ground state of a quantum system by minimising the energy of states of a particular form (``variational ansatz'')~\cite{peruzzo14}. The hope is that by optimising over a wider family of states than are accessible to classical methods, one can produce more accurate approximations to ground states. VQE may be more suitable for near-term quantum computers than some other quantum algorithms, because it uses low-depth quantum circuits which are optimised to make best use of the hardware. However, current quantum hardware is as yet unable to run VQE on instances beyond the capacity of classical exact diagonalisation.

In the period before VQE or other algorithms are able to produce ground states accurately directly, it is of interest to develop ways in which VQE can improve -- and be improved by -- existing classical methods. Here we consider the case of variational Monte Carlo (VMC) methods~\cite{sorella05,yokoyama87,giamarchi91,carleo17}. These methods allow the calculation of properties of quantum states that are too large to represent directly on a classical computer, but which nevertheless have an efficient classical description, in the sense that any given amplitude of the state can be computed efficiently (up to an overall normalisation constant).

The intent behind VMC is to approximate the true ground state $\ket{\psi_G}$ of a quantum system with such a classically efficient state $\ket{\widetilde{\psi_G}}$, and then to compute properties of $\ket{\widetilde{\psi_G}}$ in the hope that these give insight into properties of $\ket{\psi_G}$. Although, from empirical and complexity-theoretic arguments, it is not expected that all quantum ground states have an efficient description of this form, in some cases VMC methods have nevertheless been very effective~\cite{yokoyama87,giamarchi91,carleo17}.

Here we propose an approach which we call quantum-enhanced variational Monte Carlo (QEVMC), which is based on using samples produced by VQE -- or some other quantum algorithm -- as input to VMC. We show using numerical simulations and experimental data from a quantum computer that this hybrid approach can be advantageous, in three respects:
\begin{enumerate}
    \item Convergence to the target distribution can be accelerated using VQE samples, compared with a standard classically samplable initial distribution;
    \item VMC can reduce the energy compared with the energy of the original VQE state;
    \item VQE samples generated on a small quantum computer can be used within VMC to accelerate convergence for systems beyond the quantum computer's capacity.
\end{enumerate}
We validate these three points for two well-studied proof-of-principle systems: the Fermi-Hubbard model using the Gutzwiller wavefunction~\cite{gutzwiller63}, and the Ising model with transverse field using the neural network quantum state ansatz~\cite{carleo17}.

The first point above addresses a practical issue faced by VMC: convergence to the target distribution can be slow~\cite{choo20,zhao21}, which is a particular challenge given the inherently sequential nature of the MCMC process. One reason for slow convergence is that the initial distribution in VMC may be far from the target distribution. VQE is expected to be able to produce initial distributions which are closer to the target distribution, by better representing the desired quantum correlations.

Regarding the second point above, note that in a situation where VMC improves the VQE energy, we could instead have used a classical approach throughout to achieve this improved energy, without the need for the quantum computer. Thus, improving the VQE energy using VMC as a goal in itself seems to be primarily of theoretical, rather than practical, interest. However, connecting to the first point, using the quantum computer could speed up convergence to this target energy, and hence make a previously infeasible VMC procedure feasible.

\begin{table*}[t]
    \centering
    \begin{tabular}[t]{|>{\centering}p{5cm}|>{\centering\arraybackslash}p{2cm}|>{\centering\arraybackslash}p{2cm}|>{\centering\arraybackslash}p{3cm}|>{\centering\arraybackslash}p{3cm}|}
        \hline Method & Type of Hamiltonian & Needs only offline access & Type of access needed & Converges to true ground state\\
        \hline Quantum-enhanced MCMC~\cite{layden22} & Classical & \xmark & Sampling & \cmark \\
        QC-QMC~\cite{huggins22,wan23} & Quantum & \cmark & Estimating overlaps & \cmark \\
        CQA/QAEE~\cite{xu22} & Quantum & \xmark & Estimating 
        overlaps & \cmark \\
        QC-FCIQMC~\cite{zhang22} & Quantum & \xmark & Estimating elements of a Hamiltonian & \cmark \\
        VQNHE~\cite{zhang22a} & Quantum & \xmark & Sampling (rel.\ error) & \xmark\\
        NEM~\cite{bennewitz22} & Quantum & \cmark & Sampling & \xmark\\
        Data-enhanced VMC~\cite{czischek22,moss23} & Experimental & \cmark & Sampling & \xmark\\
        \textbf{Quantum-enhanced VMC} & Quantum & \cmark & Sampling & \xmark\\
        \hline
    \end{tabular}
    \caption{QEVMC compared with related work for producing low-energy states of quantum/classical systems using a hybrid quantum/classical approach. Methods which converge to the true ground state in general require exponential time to achieve this. Using QC-QMC without interactive access to the quantum computer initially required exponential processing time~\cite{huggins22}; this was improved to polynomial time in~\cite{wan23}.}
    \label{tab:comparison}
\end{table*}

A crucial advantage of QEVMC is that it can be implemented using a very simple protocol. A set of VQE samples, once generated, are stored and used to generate a distribution which can be sampled from at the start of the VMC procedure; there is no need to interact with the quantum computer after this point. In addition, we expect QEVMC to be resilient to errors in the quantum state. For example, in the simple case where the final state produced by VQE is a mixed state $\rho_V = (1-\epsilon)\psi_V + \epsilon \mu$, where $\ket{\psi_V}$ is the intended final state and $\mu$ is the maximally mixed state, the use of VMC on $\rho_V$ is equivalent to using $\ket{\psi_V}$ as input, except for an $\epsilon$ fraction of the samples, where the uniform distribution is used. The final output of QEVMC should therefore gracefully interpolate between the output generated with a perfect input state, and the output generated with the classical uniform distribution as input.

It is even possible that in some cases, a small amount of noise could be beneficial to QEVMC, as it could reduce the distance between the initial distribution generated by VQE and the target distribution.

\subsection{Related work}

Quite a number of recent works have explored ways in which near-term quantum algorithms could accelerate classical approaches for finding low-energy states of quantum (or classical) systems based on Monte Carlo methods. For a very recent survey, see~\cite{mazzola23}. We summarise these and how they compare with QEVMC in Table \ref{tab:comparison}.

A standard approach for sampling from a desired probability distribution is the Markov chain Monte Carlo method known as the Metropolis-Hastings algorithm. This algorithm produces samples from a desired probability distribution by repeatedly sampling from a proposal distribution that depends on the current state, and accepting or rejecting the sample based on a certain rule. Layden et al.~\cite{layden22} propose replacing these samples with quantum samples, where $P(x \to y) = |\braket{y|U|x}|^2$ for some symmetric unitary $U$. Based on numerical and experimental results, and theoretical intuition, they argue that this should lead to an algorithm which mixes more rapidly to the target distribution than classical proposals. Layden et al.\ apply their algorithm to producing samples from the Boltzmann distribution of classical Ising models. Attractive aspects of their approach are that it requires relatively simple quantum circuits $U$ and should be quite robust to errors experienced by the quantum computer. However, their method requires many rounds of communication with the quantum computer, as each new step of the algorithm requires $U$ to be applied to a different computational basis state. By contrast, the method proposed here relies only on the use of one set of samples from the quantum computer, which can even be produced ``offline'' in advance.

A central concept in quantum Monte Carlo methods in general is the trial state $\ket{\psi_T}$, which is intended to approximate the ground state $\ket{\psi_G}$ at the end of the algorithm. In VMC, the parameters of $\ket{\psi_T}$ are chosen using an optimisation algorithm, whereas in projector quantum Monte Carlo, the trial state is intended to approximate the state corresponding to time-evolving the initial state for various imaginary times $\tau$.

Huggins et al.~\cite{huggins22} proposed to use a quantum computer to generate trial states in projector QMC from an ansatz not accessible classically. Using these trial states within the QMC framework requires approximately computing amplitudes using a quantum computer, to small relative error. Achieving a sufficiently small error may require a large, and even exponential, number of measurements~\cite{mazzola22}, although this question is delicate and may depend on which variant of QMC is used~\cite{lee22response}. The approach of~\cite{huggins22} initially required either the use of a quantum computer iteratively at each stage of the algorithm, or exponentially costly classical processing, but the classical processing cost with offline access to the quantum computer was improved to polynomial by Wan et al.~\cite{wan23}. Xu and Li~\cite{xu22} presented a related approach, also based on quantum trial states, with an explicit application to accelerating VMC. Their approach also requires approximate quantum computation of amplitudes, and needs a new quantum circuit to be executed for each trial state.

Y.\ Zhang et al.~\cite{zhang22} developed an alternative approach to accelerating the variant of projector quantum Monte Carlo known as FCIQMC. A key issue afflicting this method is the presence of the sign problem (``non-stoquasticity'' of the Hamiltonian; a stoquastic Hamiltonian is one where all off-diagonal entries are real and non-positive). Y.\ Zhang et al.\ suggest the use of a quantum circuit $U$ produced using VQE to approximately diagonalise the Hamiltonian, such that the sign problem is mitigated. Their method requires the use of $U$ at each step of the QMC algorithm to estimate entries of the transformed Hamiltonian.

S.-X.\ Zhang et al.~\cite{zhang22a} proposed the use of VQE combined with a neural network, where the neural network is used to effectively implement a nonunitary operation acting on the state produced via VQE, and expectations of desired operators are estimated via a similar approach to that used in VMC. Their method jointly optimises over the VQE parameters and the neural network parameters, involves modifying the quantum circuit to be executed compared with the original VQE circuit, and seems to require estimating probabilities up to small relative error, in order to accurately compute a ratio of terms.

Bennewitz et al.~\cite{bennewitz22} developed the concept of ``neural error mitigation'' (NEM), whereby a neural quantum state (NQS) ansatz is used to improve the energy produced by a quantum circuit via VQE. Similarly to quantum-enhanced VMC, their approach requires only samples from measurements on the quantum state that can be stored offline. Their method is based on training the weights in the NQS such that the corresponding distribution matches the distribution produced via VQE, as far as possible, as an initial step before continuing the VMC optimisation process. This is unlike the approach proposed here, where the distribution produced by VQE is used directly as input to an MCMC algorithm. Indeed, in some cases, NQS sampling can be performed directly, without the need for MCMC.

A similar approach of pre-training NQS weights was used in the context of data from a Rydberg atom experiment (as opposed to data generated by a quantum computer running VQE), by Czischek et al.~\cite{czischek22} and Moss et al.~\cite{moss23}. These works show that training NQS weights based on experimental data can reduce convergence time in VMC optimisation (as opposed to reducing MCMC convergence time, which is our primary focus here).

Finally, Yang, Lu and Li~\cite{yang21} have proposed a hybrid quantum Monte Carlo algorithm for simulating time-dynamics with lower quantum circuit complexity than standard quantum algorithmic techniques.

\section{Variational Monte Carlo}
\label{sec:vmc}

We begin by reviewing the VMC approach. Imagine that we would like to find the ground state of a Hamiltonian $H$ on $n$ qubits. The VMC framework assumes efficient access to the following:
\begin{enumerate}
    \item The ability to sample from an initial distribution $\mathcal{D}_0$ on $\{0,1\}^n$;
    \item A mixer (Markov chain on $\{0,1\}^n$);
    \item The ability to compute amplitudes $\braket{x|\psi_{\bm\theta}}$, perhaps up to an overall normalising constant. Here $\ket{\psi_{\bm\theta}}$ is one of a family of states  parametrised by a vector~$\bm\theta$.
\end{enumerate}
We can use these tools to sample from the distribution $\mathcal{D}_{\bm\theta}$, where $\Pr_{\mathcal{D}_{\bm\theta}}[x] = |\braket{x|\psi_{\bm\theta}}|^2 =: p_{\bm\theta}(x)$, via the Metropolis-Hastings algorithm. This algorithm proceeds as follows:
\begin{enumerate}
    \item Sample $x \sim \mathcal{D}_0$.
    \item Pick a new configuration $y \in \{0,1\}^n$ from a \emph{proposal distribution} $P$.
    \item Accept the new configuration with probability
    \[ \min\left\{ 1, \frac{p_{\bm\theta}(y) P(y \to x)}{p_{\bm\theta}(x) P(x \to y)} \right\}. \]
    \item Go to step 2.
\end{enumerate}
One can prove that (under weak conditions on $P$) eventually this algorithm produces samples from the distribution $p_{\bm\theta}$ up to arbitrary precision. This procedure has two downsides:
\begin{itemize}
    \item Samples following the first sample are correlated, with samples taken at close times experiencing higher levels of correlation. To avoid this issue, one can discard samples in between the first sample and subsequent ones retained.
    \item It can take many steps of the random walk to converge to the desired distribution $\mathcal{D}_{\bm\theta}$. This is particularly likely to be an issue if $\mathcal{D}_{\bm\theta}$ is far away from~$\mathcal{D}_0$.
\end{itemize}

Note that all we needed to execute the algorithm is the ability to produce samples from $\mathcal{D}_0$, and evaluate probabilities $p_{\bm\theta}(x)$ up to an overall scaling constant. In particular, we did not need to be able to sample from $\mathcal{D}_{\bm\theta}$ (which is what we were aiming to do), or to evaluate the normalising constant $\sum_x p_{\bm\theta}(x)$.

Once we have samples from $\mathcal{D}_{\bm\theta}$, we can compute observables $\braket{\psi_{\bm\theta}|M|\psi_{\bm\theta}}$, using
\begin{align*}
    \braket{\psi_{\bm\theta}|M|\psi_{\bm\theta}} &= \sum_{x \in \{0,1\}^n} \braket{\psi_{\bm\theta}|M|x}\braket{x|\psi_{\bm\theta}}\\
    &= \sum_{x\in\{0,1\}^n} \frac{\braket{\psi_{\bm\theta}|M|x}}{\braket{\psi_{\bm\theta}|x}} |\braket{x|\psi_{\bm\theta}}|^2.
\end{align*}
The quantity $\frac{\braket{\psi_{\bm\theta}|M|x}}{\braket{\psi_{\bm\theta}|x}}$, called the local energy in the case where $M=H$, can be computed efficiently for many operators $M$ (e.g.\ if $M$ is a sum of a small number of Pauli matrices). Thus, averaging the local energy over samples from $\mathcal{D}_{\bm\theta}$ gives an approximation of $\braket{\psi_{\bm\theta}|M|\psi_{\bm\theta}}$. Given that this expression is a ratio of two terms involving $\ket{\psi_{\bm\theta}}$, $\ket{\psi_{\bm\theta}}$ does not need to be normalised.

The overall VMC framework then optimises over variational parameters $\bm\theta$ to find a state $\ket{\psi_{\bm\theta}}$ whose energy with respect to $H$ is minimised.

\section{Quantum-enhanced variational Monte Carlo}

In VMC, the initial distribution $\mathcal{D}_0$ is taken to be a ``simple'' distribution which is related to the desired ground state. For example, a uniform distribution (for a spin system), or a Slater determinant (for a fermionic system). Here we propose an approach which we call quantum-enhanced variational Monte Carlo (QEVMC), where these initial samples are replaced with samples from a quantum state prepared using VQE. The rest of the VMC algorithm remains unchanged. In particular, these initial samples can be produced by an experiment using a quantum computer, and then stored offline until they are used in VMC.

The intuition for why this could be advantageous is as follows. The quantum state generated using VQE should have a higher fidelity with the true ground state of $H$ than the initial state used in VMC. Therefore, it should take fewer steps to reach the ground state distribution (if one could compute its amplitudes) than starting with the VMC initial state. In VMC, we do not aim to reach the true ground state, but a (variational) approximation. But if this approximation is close to the true ground state, achieving high fidelity with the ground state implies high fidelity with the approximation.

One can make this argument more rigorous as follows. Define the $\chi^2$-divergence between distributions $\nu$ and $\pi$ as
\[ \chi^2(\nu,\pi) = \sum_x \pi(x) \left( \frac{\nu(x)}{\pi(x)}-1 \right)^2 = \sum_x \frac{\nu(x)^2}{\pi(x)} - 1. \]
Let $C$ be an ergodic Markov chain with stationary distribution $\pi$. Then, given a sequence of distributions $\nu_0,\dots,\nu_n$, where $\nu_{i+1}$ is produced by applying one step of $C$ to $\nu_i$, we have~\cite{fill91}
\[ 4\|\nu_n - \pi\|_1^2 \le \lambda_2(M)^n \chi^2(\nu_0,\pi), \]
where $\lambda_2(C) < 1$ denotes the second largest eigenvalue of $C$. This implies that, in order to achieve $\|\nu_n - \pi\|_1 \le \epsilon$, it is sufficient to take
\[ n = O\left(\frac{\log(\chi^2(\nu_0,\pi)/\epsilon)}{\log(1/\lambda_2(C))} \right) = O\left(\tau \log(\chi^2(\nu_0,\pi)/\epsilon)\right), \]
where $\tau = 1/(1 - \lambda_2(C))$ is the relaxation time of $M$. This is not straightforward to compute as it depends on the probability of acceptance in the Metropolis-Hastings algorithm.

The $\chi^2$-divergence can be related to the fidelity between the corresponding quantum states $\ket{\nu}$ and $\ket{\pi}$ obtained by taking the square roots of the probabilities via Jensen's inequality, which shows that
\[ \frac{1}{\sqrt{\chi^2(\nu,\pi) + 1}} \le \sum_x \nu(x) \sqrt{\frac{|\pi(x)|}{|\nu(x)|}} = \braket{\nu|\pi}  \]
and hence $\chi^2(\nu,\pi) \ge 1/\braket{\nu|\pi}^2 - 1$. This is a lower bound (as opposed to an upper bound), and only holds if $\ket{\nu}$ and $\ket{\pi}$ have the same phase for every amplitude, but it is still suggestive: in order to obtain a small $\chi^2$-divergence, one needs to have a high fidelity. On the other hand, if the fidelity between the starting state and the desired final distribution is small (for example, exponentially small) we will obtain a high $\chi^2$-divergence and expect a long mixing time.

One situation where an initial distribution $\nu_0^{VQE}$ produced by VQE could outperform an initial distribution $\nu_0^{C}$ produced classically is where $\chi^2(\nu_0^{VQE},\pi) = 2^{\alpha n}$, and $\chi^2(\nu_0^{C},\pi) = 2^{\beta n}$, for some $0 < \alpha < \beta$. Then we would expect a speedup by a factor of $\sim \beta/\alpha$. A more significant speedup would be obtained if $\chi^2(\nu_0^{VQE},\pi) = O(1)$ -- corresponding to the VQE state having $\Omega(1)$ fidelity with the VMC state -- where the expected speedup would be a factor of $\Theta(n)$. These theoretical bounds only provide intuition; the speedups achieved in practice could be better (or worse).

\section{Results}

We apply QEVMC to two well-studied models in condensed matter physics: the Fermi-Hubbard model, using the Gutzwiller wave function ansatz~\cite{yokoyama87}, and the Ising model with transverse field, using neural network quantum states~\cite{carleo17}.

\subsection{Fermi-Hubbard model}

We aim to find the ground state of instances of the Fermi-Hubbard model,
\[ H = -\sum_{\braket{i,j},\sigma} a^\dag_{i\sigma} a_{j\sigma} + a_{j\sigma}^\dag a_{i\sigma} + U \sum_k n_{k\uparrow} n_{k\downarrow}, \]
where $\sigma \in \{\uparrow,\downarrow\}$ and the sum is over sites $(i,j)$ in a rectangular lattice. This model has been extensively used as a benchmark for both classical and quantum variational methods~\cite{yokoyama87,Cade2020,Stanisic2021}.

A variational ansatz which is frequently used in VMC is the Gutzwiller wave function~\cite{gutzwiller63},
\[ \ket{\psi_G} \propto e^{-c \sum_k n_{k\uparrow} n_{k\downarrow} } \ket{\psi_{SD}}, \]
where $\ket{\psi_{SD}}$ is a Slater determinant corresponding to the ground state of the hopping part of $H$, and $c>0$. The intention behind this ansatz is to penalise basis states with high double occupancy, which will receive a large energy penalty from the onsite part of $H$. Amplitudes of $\ket{\psi_G}$ can be evaluated efficiently classically, and to produce this wave function in the VMC framework, it is natural to use as initial state the Slater determinant $\ket{\psi_{SD}}$, whose corresponding distribution can be sampled from directly. As a mixer, one can use the random walk which moves a randomly chosen electron from one site to a neighbouring site~\cite{yokoyama87}. Computationally, this corresponds to updating the bitstring by flipping a neighbouring pair of bits from 01 to 10 (or vice versa) within one spin sector.

Here we initially applied QEVMC to the Gutzwiller wavefunction for system sizes that are small enough to be simulated exactly classically, and which are also accessible to quantum computing experiments. This enabled us to carry out a full analysis and comparison of the performance of the algorithm.

We applied QEVMC to experimental data that had already been generated in experiments applying VQE to the Fermi-Hubbard model on quantum hardware~\cite{Stanisic2021}. These experiments optimised over the family of quantum circuits known as the Hamiltonian variational ansatz~\cite{Wecker2015} to find good circuits for preparing approximate ground states of the Fermi-Hubbard model. Although in some cases the fidelity of the state produced with the true ground state was relatively low, qualitative physical features of the ground state were nevertheless reproduced~\cite{Stanisic2021}. In this work we focus on three examples from~\cite{Stanisic2021}: the $1\times 4$ Fermi-Hubbard model with two variational layers; the $1 \times 8$ Fermi-Hubbard model with one variational layer; and the $2 \times 4$ Fermi-Hubbard model with one variational layer. These are all instances where more variational layers would be required to achieve a high-fidelity approximation to the ground state. In all cases we take $U=4$ and work at half-filling.

First, we looked at the accuracy of the algorithm as it proceeds, for a particular choice of Gutzwiller parameter $c$ which approximately minimises the energy ($c \approx 0.421$ for $1\times 4$, $c \approx 0.431$ for $1\times 8$, $c \approx 0.453$ for $2 \times 4$). We measured the accuracy via the total variation distance from the desired probability distribution, and the error in the energy.

The initial distributions that we compared were:
\begin{itemize}
    \item Slater determinant ($U=0$ ground state), the standard starting distribution for this method;
    \item Theoretical VQE distributions obtained via classical simulation;
    \item Experimentally obtained distributions, taking advantage of error mitigation via postselection on occupation number, and averaging over time-reversal and reflection symmetries of the Fermi-Hubbard model~\cite{Stanisic2021}.
\end{itemize}

\begin{figure*}[h!]
    \centering
    \includegraphics[width=0.3\textwidth]{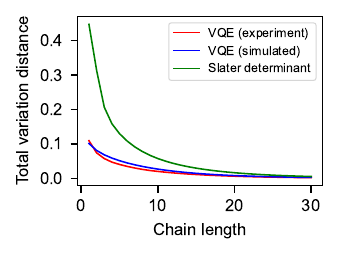}
    \includegraphics[width=0.3\textwidth]{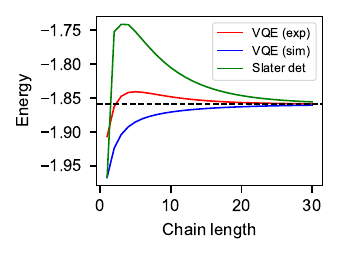}
    \includegraphics[width=0.3\textwidth]{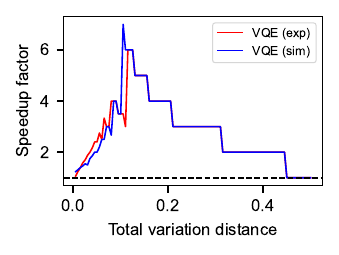}
    \caption{QEVMC applied to a $1\times 4$ Fermi-Hubbard instance at half-filling. Speedup factor is the ratio between the chain length required to achieve a given total variation distance from the target distribution when starting with a Slater determinant, vs.\ starting with a VQE distribution. VQE distributions are taken from~\cite{Stanisic2021}.}
    \label{fig:fermihubbard1x4}
\end{figure*}

\begin{figure*}[h!]
    \centering
    \includegraphics[width=0.3\textwidth]{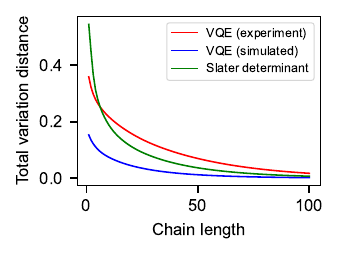}
    \includegraphics[width=0.3\textwidth]{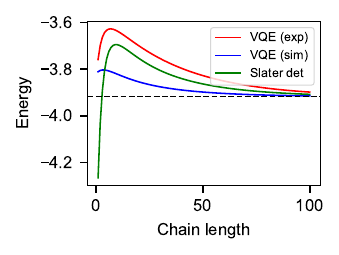}
    \includegraphics[width=0.3\textwidth]{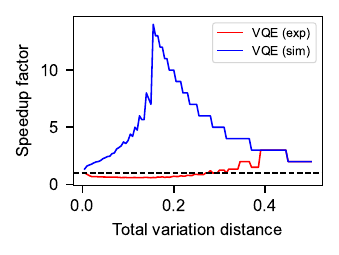}
    \caption{QEVMC applied to a $1\times 8$ Fermi-Hubbard instance at half-filling. VQE distribution taken from~\cite{Stanisic2021}.}
    \label{fig:fermihubbard1x8}
\end{figure*}

\begin{figure*}[h!]
    \centering
    \includegraphics[width=0.3\textwidth]{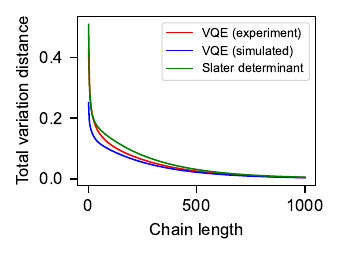}
    \includegraphics[width=0.3\textwidth]{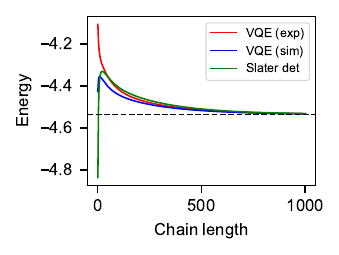}
    \includegraphics[width=0.3\textwidth]{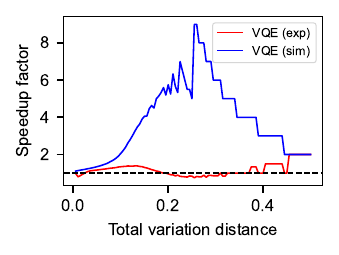}
    \caption{QEVMC applied to a $2\times 4$ Fermi-Hubbard instance at half-filling. VQE distribution taken from~\cite{Stanisic2021}.}
    \label{fig:fermihubbard2x4}
\end{figure*}

The results are shown in Figures \ref{fig:fermihubbard1x4}, \ref{fig:fermihubbard1x8} and \ref{fig:fermihubbard2x4}. One can see that in all cases, starting with the VQE distribution does lead to convergence to the desired distribution. Also, in the cases of the $1\times 4$ experimental results, and all of the simulated results, the distributions obtained by starting with the VQE distribution have a lower total variation distance throughout the algorithm. However, the $1\times 8$ experimental results do not achieve a lower total variational distance throughout the whole algorithm. In the $1\times 8$ and $2\times 4$ cases, the final energies obtained by VMC are lower than the energies achieved by the VQE distribution alone (which are $-3.37$ and $-3.71$, respectively, for the experimental distributions~\cite{Stanisic2021}). Note that the local energy reported at the start of the VMC algorithm is not necessarily an accurate reflection of the energy achieved by the VQE distribution.

We additionally measured the speedup that would be obtained via VQE, by measuring the number of steps of the Metropolis algorithm that are required to achieve a given total variation distance level, when starting with a Slater determinant, vs the VQE initial distribution. The results are shown in terms of the ratio between the number of steps taken by QEVMC, vs.\ the number of steps taken when starting with a Slater determinant. Here we found that the simulated VQE distributions achieve a significant speedup (e.g.\ up to $14\times$ for $1\times 8$).  The $1\times 4$ experimental distribution achieves a speedup up to $6\times$, but there is little speedup obtained in the $1\times 8$ and $2\times 4$ experiments.

In the above experiments, in order to exactly determine the performance of the various VMC starting distributions, we directly implemented the corresponding stochastic matrix multiplication, which is inherently inefficient for large problem sizes. We additionally implemented the true VMC method as described in Section \ref{sec:vmc}. In addition to providing a check on the performance of the algorithm in practice, this enabled us to use the VQE distribution within VMC for problem sizes beyond the capacity of classical exact diagonalisation.

\begin{figure*}[h!]
    \centering
    \begin{subfigure}[b]{0.3\textwidth}
    \centering
    \includegraphics[width=\textwidth]{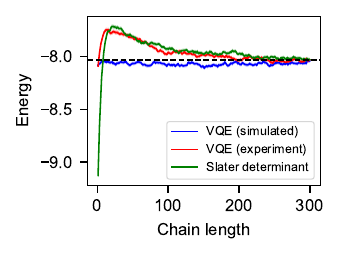}
    \caption{$1\times 16$}
    \end{subfigure}
    \begin{subfigure}[b]{0.3\textwidth}
    \centering
    \includegraphics[width=\textwidth]{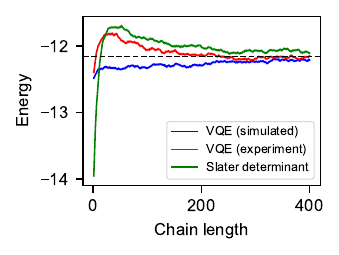}
    \caption{$1\times 24$}
    \end{subfigure}
    \begin{subfigure}[b]{0.3\textwidth}
    \centering
    \includegraphics[width=\textwidth]{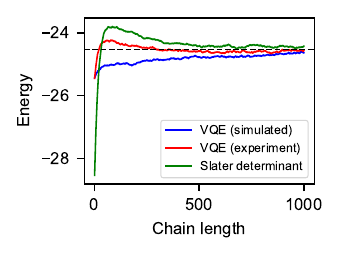}
    \caption{$1\times 48$}
    \end{subfigure}
    \caption{Using small VQE distributions as starting distributions for larger VMC runs. Solid lines represent the mean energy obtained over 10,000 runs, shaded regions are standard error of the mean. Dashed horizontal lines are estimates of the target energy obtained from a VMC run with 1M steps starting from a Slater determinant.}
    \label{fig:fhlargesizes}
\end{figure*}

To demonstrate this, we considered $1\times L$ Fermi-Hubbard instances for $L \in \{16,24,48\}$, corresponding to problems on up to 96 qubits. We use the VQE distribution for $1\times 8$ Fermi-Hubbard to generate initial samples from this distribution by concatenating $L/8$ independent samples, and aim to sample from a distribution corresponding to a state with the same Gutzwiller parameter as in $1\times 8$ ($c \approx 0.431$). The results are shown in Figure \ref{fig:fhlargesizes}. One can see that even for significantly larger problem sizes than are accessible directly on the quantum computer, use of the VQE initial distribution enables convergence to the target energy within a reasonable number of steps, and in some cases apparently faster than using a Slater determinant. Intriguingly, in the case of the $1\times 48$ system, use of the experimental VQE distribution seems to give the fastest convergence. However, given that the total variation distance to the desired distribution is not accessible for large system sizes, it is unclear whether this effect would persist across a wide range of other observables.

\subsection{Ising model with transverse field}

Here we aim to find the ground state of the 1D Ising model with transverse field (TFI),
\[ H = - J \sum_{i=1}^{N-1} Z_i Z_{i+1} - h \sum_{j=1}^{N} X_j, \]
where $N$ is the number of lattice sites, $J=1$, and we assume open boundary conditions (OBC) unless specified otherwise.

A successful ansatz for this model is the neural network quantum state (NQS) ansatz proposed by Carleo and Troyer~\cite{carleo17} and given as
\[
\ket{\psi(\mathcal{W})} \propto \sum_{S} \Psi(S; \mathcal{W}) \ket{S}
\]
where $\ket{S} = \ket{s_1}\dots\ket{s_N}$ is a computational basis state, $S$ is the bitstring which denotes that basis state, and the coefficients are defined as
\[
\Psi(S; \mathcal{W}) = \sum_{\{h_i\}} \mathrm{exp} \left[ \sum_{j=1}^{N} a_j \sigma_j^z + \sum_{i=1}^{M} h_i (b_i + \sum_{j=1}^{N} W_{ij} \sigma_j^z )\right],
\]
where $\sigma_j^z = \braket{S|Z_j|S}$, and $h_i = \{-1, 1\}$ corresponds to $M$ hidden variables.
The set of weights $\mathcal{W} = \{a_i, b_j, W_{ij}\}$ are the parameters we optimize to find the closest NQS to the ground state of some given Hamiltonian.
As in the original paper~\cite{carleo17}, we are adding only a single hidden layer to the NQS, therefore we use the traced out expression giving the simplified equation for any bitstring $S$,
\[
\Psi(\mathcal{S}; \mathcal{W}) =   \mathrm{exp} \left[ \sum_{j=1}^{N} a_j \sigma_j^z \right] \times \prod_{i=1}^{M} F_i(S),
\]
where $F_i(S) = 2 \cosh \left[b_i+\sum_{j=1}^{N} W_{ij} \sigma_j^z \right]$.
While generally the weights for the NQS ansatz are complex-valued, motivated by the solutions for the TFI model on 40 and 80 qubits in~\cite{carleo17} which are real-valued, we take them to be real-valued, halving the parameter search space on this problem.
We examine three regimes. Taking $J=1$, we have $h \in \{0.5, 1.0, 2.0\}$, with $J/h=1$ being the critical phase-transition point, and we use the number of nodes in the hidden layer to be $M=\alpha N$ with $\alpha \in \{1, 2, 4\}$.
The data presented in the paper here is for $h=1$ unless otherwise specified, however the results for $h=0.5$ and $h=2$ were not qualitatively different.

To find the NQS that best describes the ground state of TFI model, the algorithm requires us to first run an optimization over the NQS weights ($\mathcal{W}$), minimizing the expected energy of the NQS with respect to the TFI Hamiltonian.
The standard algorithm used for this optimization is the stochastic reconfiguration (SR) algorithm~\cite{Sorella1998, Sorella2007}.
The initial NQS for this optimization is chosen by selecting random weights.
Then MCMC is used to acquire the necessary sample distribution and calculate the expected energy values of the NQS at each iteration of the optimization algorithm.
The initial distribution is taken to be a uniform distribution, and single spin flip is used as a mixer.
This corresponds to updating the bitstring by flipping a single spin chosen at random from $0$ to $1$ (or vice versa).
Each bitstring is updated independently (this corresponds to running the MC algorithm many times from different initial bitstrings randomly chosen from the initial distribution to acquire the bitstring from the final distribution).
Given this final distribution of samples at each iteration of SR, to calculate the expected energy we calculate the local energy of the samples.
This is given as
\[
E_{\mathrm{loc}}(S) = \frac{\braket{S|H|\Psi(\mathcal{W})}} {{\Psi(S; \mathcal{W}) }},
\]
where $\Psi(\mathcal{W})$ is the full wavefunction NQS describes, and $S$ is a given bitstring sample from the final distribution.
We then take the average of these local energy estimates for the final energy estimate.

We investigate here a potential speed-up using an already known VQE distribution as the initial distribution for the Monte Carlo procedure.
The impact of this could be two-fold. For an already known NQS which describes the ground state of a Hamiltonian, we could speed up the mixing time from the initial distribution to the NQS distribution.
If the NQS is not known, the overall number of iterations the SR procedure needs to take to a given error could offer a speed-up, therefore needing a smaller number of samples overall.

To generate the VQE distribution we use the Hamiltonian Variational ansatz using an all plus state as the initial state.
To generate the Hamiltonian Variational ansatz, we split up the TFI Hamiltonian into even $ZZ$ terms, odd $ZZ$ terms, and the the single qubit transverse field terms.
In a single layer of the ansatz we first apply all the even term gates parameterized by a single parameter, then odd terms parameterized by another parameter, and finally all the field terms which have a third parameter.
Therefore each layer of VQE uses three parameters, and we find optimal VQE states using the BFGS algorithm for a set number of layers.
We examine solutions up to four layers as we would like the states produced by VQE not to approximate the ground state too well.

We consider a few different system sizes, similarly to the Fermi-Hubbard model.
We examine various aspects of the speed-up for a $16$ spin system via exact calculations and we then examine a $24$ spin system with the true VMC method.
Finally, we also look at $L \in \{40, 80\}$ spin system using the NQS that was found in the original work~\cite{carleo17} as approximate ground states of the TFI model to examine potential speed-up by concatenating $L/20$ independent samples from the VQE distribution for a $20$-spin system.

To examine potential speed-up the VQE distribution gives to the MCMC for a known NQS, we first find good NQS states using the stochastic reconfiguration algorithm with a fixed number of iterations. 
We use 200 iterations for $h \in \{0.5, 1.0, 2.0\}$. 
At each iteration, to estimate the energy of the NQS at the current iteration, we use $5000$ samples from the initial distribution (here, uniform) and take the last state of the Markov chain of length $15$/$20$ for system size $16$/$24$.
We estimate the final energy by using the average of local energy estimate from the last five iteration, getting the relative energy errors wrt to the ground state energy to be less than $0.066/0.0042/7.4 \times 10^{-7}$ for $h \in \{0.5, 1.0, 2.0\}$ demonstrating good agreement with the ground state energy.

\begin{figure*}[t]
    \centering
    \includegraphics[width=0.3\textwidth]{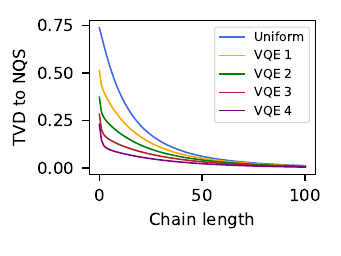}
    \includegraphics[width=0.3\textwidth]{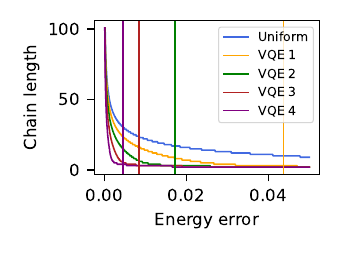}
    \includegraphics[width=0.3\textwidth]{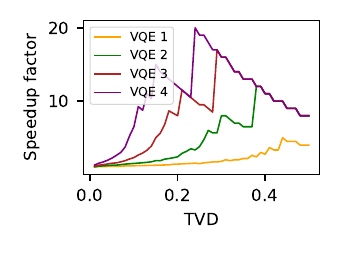}
    \caption{QEVMC applied to a $16$ spin TFI instance ($h=1$) using an already trained NQS ($\alpha=1$). The values presented here are calculated using exact distributions. Initial distributions used marked in different colours, ``Uniform'' stands for uniform distribution, ``VQE $i$'' for VQE distribution obtained from the Hamiltonian Variational Ansatz with $i$ layers and obtaining samples from the final state in the computational basis. The second plot has vertical lines showing the relative energy error of the VQE state, calculated exactly (color coded as on the legend). We can notice that these energy errors are higher than the final errors after running VMC using the VQE distributions.
    }
    \label{fig:TFI16_endpoint}
\end{figure*}

As we now have a good representation of the ground state as an NQS, we can inspect the difference in mixing time between VQE and the uniform distribution, using a similar approach to the small Fermi-Hubbard instances considered above.
For the $16$ spin system, we can calculate the exact total variation distance (TVD) of the proposed distribution to the desired NQS distribution as a function of the number of steps of the Markov chain.
We can also calculate the length of the Markov chain needed to achieve a specific relative energy error (wrt the NQS energy) as well as to achieve a specific TVD (to the NQS distribution).
We then compare performance of these different distributions, by comparing the length of MC chain needed to achieve a specific TVD taking the uniform distribution as the baseline.
We can see some of the results in Figure~\ref{fig:TFI16_endpoint}. 

\begin{figure*}[t]
    \centering
    \includegraphics[width=0.3\textwidth]{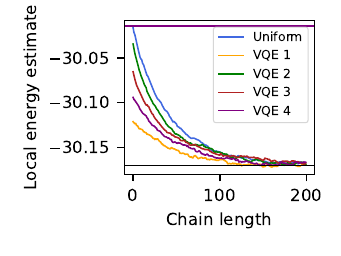}
    \includegraphics[width=0.3\textwidth]{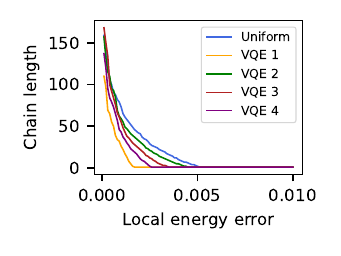}
    \includegraphics[width=0.3\textwidth]{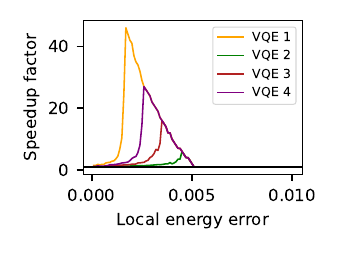}
    \caption{QEVMC applied to a $24$ spin TFI instance ($h=1$) using an already trained NQS ($\alpha=1$). Energy estimates were generated using $5000$ samples for each datapoint. Energy error given is relative error to the estimated NQS energy. To calculate the NQS energy estimate, a chain of length $500$ was acquired with $5000$ samples and using the uniform distribution as the starting distribution; then the mean value of the last $10$ points of the chain is taken. On the first plot we can see the NQS energy estimate as black horizontal line, we can also see the exact VQE energy for the four layer state which is much higher than the energy estimated using VMC.
    }
    \label{fig:TFI24_endpoint}
\end{figure*}

Similarly, for the $24$ spin system, we no longer easily have access to exact values, but we can look at energy estimates reached at different points in the Markov chain, and compare the errors of the distributions.
For some of the results, refer to Figure~\ref{fig:TFI24_endpoint} (for this specific result we run the SR algorithm slightly longer, for $300$ iterations and get a relative energy error with ground state energy to be 0.0017).
We can see that in the case of $16$ spin system with $h=1$, there is a speed-up of up to $20$ times when using a four layer VQE state compared to the uniform distribution.
For the $24$ spin system with $h=1$, we see a speed-up of up-to $46$ times when using a one layer VQE state compared to the uniform distribution. 


\begin{figure*}[t]
    \centering
    \includegraphics[width=0.24\textwidth]{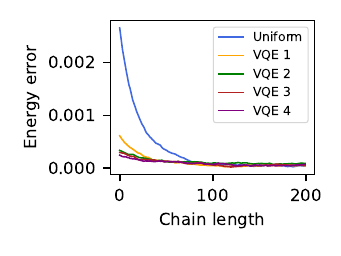}
    \includegraphics[width=0.24\textwidth]{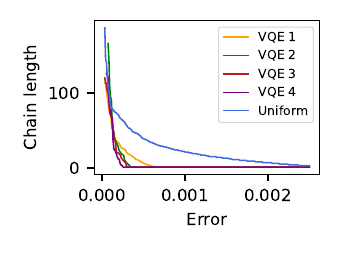}
    \includegraphics[width=0.24\textwidth]{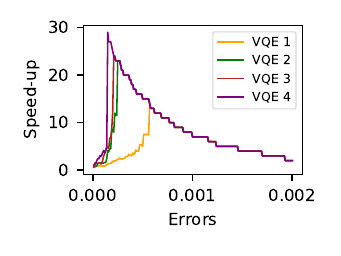}
    \includegraphics[width=0.24\textwidth]{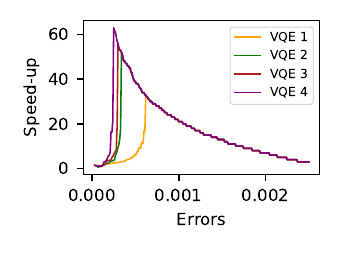}
    \caption{
        QEVMC applied to a $40$ and $80$ spin TFI instance ($h=1$) using an already trained NQS ($\alpha=4$).
        Energy error is relative error to estimated ground state energy using the data from the original paper~\cite{carleo17}.
        First figure shows the relative error as the length of MCMC increase in the $80$ spin instance, second figure shows the needed chain length to achieve a certain relative error; showing that the VQE as initial distribution allows certain error rates to be reached quicker; the last two figures show the speedup factor in terms of the needed length to a certain error when compared to the uniform distribution as initial distribution in the $40$ and $80$ spin case respectively.
        We can notice that the speedup scales up with the increase in the size of the system.
    }
    \label{fig:TFI80_endpoint}
\end{figure*}

Next, we can also examine the case of $40$- and $80$-spin systems in a similar manner to the $24$ spin system, using the NQS results from~\cite{carleo17}.
Here we use samples generated by $20$-spin systems with open boundary conditions (OBC), while the NQS from~\cite{carleo17} is for systems with periodic boundary conditions (PBC).
While one could consider running VQE for $20$-spin systems with PBC, we hypothesize that OBC are more likely to give good results due to how the two distributions are stitched together (specifically, there is no need for correlation between the first and the 20th spin).
Here, the initial samples were drawn independently from the $20$-spin VQE distribution and combined to give an initial sample for the larger system.
It is an open question whether alternative ways to piece different distributions together given the physics of the model could offer better speedups.
We look at the behaviour of the relative energy error over a chain of length $200$ with $5000$ samples taken for the energy estimate.
Results can be seen in Figure~\ref{fig:TFI80_endpoint}.
We can see from the Figure that even though the VQE distribution was acquired on a smaller system, it allows for speedups on larger systems, and the speedup scales up for a larger system size ($80$ vs $40$ spins).

\begin{figure*}[t]
    \centering
    \includegraphics[width=0.3\textwidth]{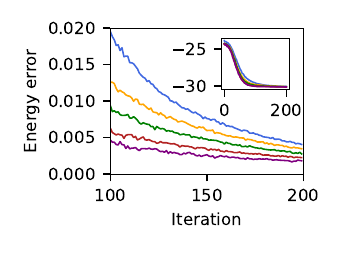}
    \includegraphics[width=0.3\textwidth]{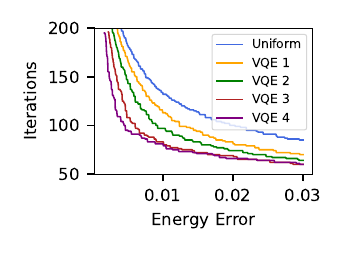}
    \includegraphics[width=0.3\textwidth]{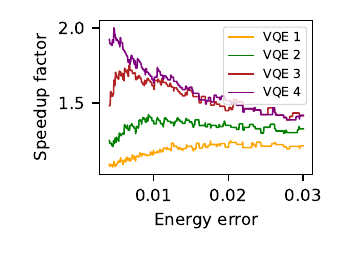}
    \caption{QEVMC applied to a $24$ spin TFI instance during stochastic reconfiguration algorithm for $h=1$. Colour coding is blue for uniform distribution, orange/green/red/purple for distribution acquired using 1/2/3/4 layers of VQE. Middle image shows SR iterations vs energy error. Right picture shows the speed-up factor of the VQE distributions in comparison to uniform. Energy error is relative error to exact ground state energy.}
    \label{fig:SR_comparison}
\end{figure*}

Finally, we will want to see if it is possible to speed-up running of the SR algorithm by using VQE distribution instead of uniform distribution as the starting distribution for the MCMC at each iteration of the SR algorithm.
We will make this comparison by comparing the number of SR iterations we need to achieve the same relative error using different distributions.
We first have to make sure that the length of the MCMC used before when running the SR algorithm is indeed suitable for these various distributions.
We visually inspect the length of chain needed for a randomly chosen NQS target distribution to be approximated given as initial distribution various VQE generated distributions compared to the uniform distribution.
We find that after $20$ to $30$ steps, there is a qualitative agreement between the distributions for the settings inspected here ($24$ qubits and $h \in \{0.5, 1.0, 2.0\}$; in the case of $16$ qubits, the number of steps is even lower).
So fixing as before the MCMC length to $20$ at each iteration of the SR algorithm (taking $5000$ samples), we find that the VQE distribution as the initial distribution to mix towards the NQS distribution gives improvement over uniform distribution in the case of $24$ qubits at various $h$, see Figure~\ref{fig:SR_comparison}, with more VQE layers allowing for higher speedup.

\subsection{Origin of the QEVMC speedup}

\begin{figure*}
    \begin{subfigure}[b]{0.4\textwidth}
    \centering
    \includegraphics[width=\textwidth]{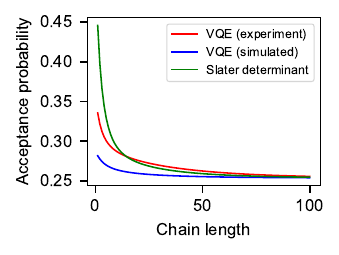}
    \caption{$1\times 8$ Fermi-Hubbard}
    \end{subfigure}
    \hspace{1cm}
    \begin{subfigure}[b]{0.4\textwidth}
    \centering
    \includegraphics[width=\textwidth]{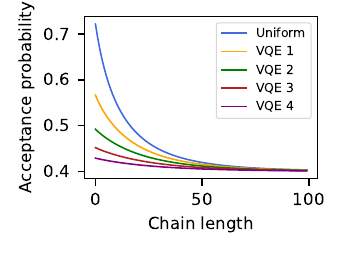}
    \caption{$1\times 16$ TFIM}
    \end{subfigure}
    \caption{Acceptance probabilities for QEVMC compared with starting with classical distributions. Labels are the same as in Figures \ref{fig:fermihubbard1x4} and \ref{fig:TFI16_endpoint}.}
    \label{fig:acceptancerate}
\end{figure*}

It is a natural question whether one can pin down the origin of the speedup achieved by QEVMC. Given that the mixer is unchanged compared with the standard Metropolis-Hastings algorithm, there are two plausible possibilities for how this speedup is achieved:
\begin{enumerate}
\item The initial distribution produced by VQE is closer to the target distribution (so fewer steps are required to reach it);
\item The acceptance rate of MCMC steps is higher when starting with the VQE distribution (so more of the steps make progress towards the target distribution).
\end{enumerate}
In Figure \ref{fig:acceptancerate} we present evidence that the second option is not the case, by showing that the average acceptance rate is lower for QEVMC for both the Fermi-Hubbard model and TFIM, albeit not substantially. This suggests that the first option is likely to be the correct explanation for the acceleration achieved by QEVMC.

\section{Discussion}

We have shown that QEVMC can outperform standard classical VMC in terms of convergence speed to the target distribution. Whether this advantage translates into a genuine speed improvement in practice depends on a number of factors, including the clock speed of the quantum computer and the behaviour of the MCMC algorithm. For example, the level of correlation considered to be acceptable in samples generated following the first sample from the target distribution affects the gap that must be left between subsequent samples that are kept, and the level of speedup available to the quantum algorithm. Chains which require a long ``burn in'' time to sample from the target distribution are the most promising candidates to achieve a significant quantum speedup.

An interesting follow-up question in the setting of our NQS results could be regarding the performance of an adaptive algorithm where the length of the chain and the number of samples changes as we get closer to the ground state.
As a marker of this proximity to the ground state, one could use the variance of the local energy estimate.
Since the VQE distribution gives an increasing boost in performance as the NQS becomes closer to the ground state description, in such an adaptive algorithm, one would expect the VQE distribution to give further speedups over the uniform distribution.


Several works have proposed the use of experimental quantum states or those generated via VQE to train weights in a NQS ansatz~\cite{bennewitz22,czischek22,moss23}, and it has been shown that this procedure can reduce the cost of VMC optimisation~\cite{czischek22,moss23}. It would be interesting to compare the performance improvements that can be obtained via this approach with the MCMC convergence speedups achieved here.

An alternative approach to the use of QEVMC is to generate the VMC target distribution directly on a quantum computer. For example, algorithms are known for constructing the Gutzwiller wave function ~\cite{murta21,seki21} whose quantum circuit depth is not substantially greater than that of preparing the Slater determinant $\ket{\psi_{SD}}$. However, these algorithms are based on the use of postselection, so may require many runs of the corresponding quantum circuit to prepare the desired state. This state could then be used as initial distribution for VMC, in a similar way to QEVMC. This could be advantageous in the realistic scenario that noise in the quantum circuit leads to errors in the state prepared. Then VMC would enable the generation of samples arbitrarily close to the target distribution.

The simple structure of the QEVMC algorithm opens up a use case for near-term quantum computers, where samples from a classically intractable distribution are produced and stored for later use as input to a classical algorithm. We expect that other applications of this concept may be found.

\subsection*{Acknowledgements}

We would like to thank M.\ Schuyler Moss for pointing out references~\cite{bennewitz22,czischek22,moss23}, Jan Lukas Bosse and Lana Mineh for their work on the VQE implementation used in this project, and the rest of the Phasecraft team for helpful comments on this work. This project has received funding from the European Research Council (ERC) under the European Union's Horizon 2020 research and innovation programme (grant agreement No.\ 817581).

\bibliographystyle{mybibstyle}
\bibliography{bibliography}

\end{document}